\documentstyle[11pt]{article}
\input pictex.sty
\font\thinlinefont=cmr5
\begin{document}
\vspace*{-2cm}
\noindent
\hspace*{11cm}
UG--FT--69/96 \\
\hspace*{11cm}
hep-ph/9703410 \\
\hspace*{11cm}
March 1997 \\
\begin{center}
\begin{large}
{\bf CP violation from new quarks in the chiral limit \\}
\end{large}
~\\
F. del Aguila, J. A. Aguilar--Saavedra \\
{\it Departamento de F\'{\i}sica Te\'{o}rica y del Cosmos,
Universidad de Granada \\
18071 Granada, Spain} \\
~\\
G. C. Branco \\
{\it Departamento de Fisica and CFIF/UTL, 
Instituto Superior T\'{e}cnico \\
1096 Lisboa, Portugal} \\
\end{center}
\begin{abstract}
We characterize CP violation in the ${\mathrm SU(2) \times U(1)}$ model due to
an extra vector-like quark or sequential family,
giving special emphasis to the chiral
limit $m_{u,d,s}=0$. In this limit, CP is conserved in the three generation
Standard Model (SM), thus implying that all CP violation is due to the two new
CP violating phases whose effects may manifest either at high energy in
processes involving the new quark or as deviations from SM unitarity equalities
among imaginary parts of invariant quartets (or, equivalently, areas of
unitarity triangles). In our analysis we use an invariant formulation,
independent of the choice of weak quark basis or the phase convention in the
generalized Cabibbo-Kobayashi-Maskawa matrix. We identify the three weak-basis
invariants, as well as the three imaginary parts of quartets $B_{1-3}$
which, in the chiral
limit, give the strength of CP violation beyond the SM. We find that for an
extra vector-like quark $|B_i| \leq 10^{-4}$, whereas for an extra sequential
family $|B_i| \leq 10^{-2}$.
\end{abstract}
\hspace*{0.8cm}
\section{Introduction}
In the Standard Model (SM) CP violation is parametrized by one CP violating
phase in the Cabibbo-Kobayashi-Maskawa (CKM) matrix \cite{papiro1}. Although
this phase accounts for the observed CP violation in the $\mathrm K^0$-$\mathrm
\bar K^0$ system \cite{papiro2}, there is no deep understanding of CP
violation. Furthermore, it has been established that the amount of CP violation
present in the SM is not sufficient to generate the baryon asymmetry of the
universe \cite{papiro3}. This provides motivation for looking for new sources
of CP violation which can lead to deviations from the SM predictions for CP
asymmetries in ${\mathrm B}^0$ decays and/or to new signals of CP violation
observable at high energy, in future colliders.

In the SM the CKM matrix $V_{\mathrm CKM}$ is a $3 \times 3$ unitary matrix
whose matrix elements $V_{ij}$ are strongly constrained by unitarity which
implies, for example, that the imaginary parts of all invariant quartets,
${\mathrm Im}\;V_{ij} V_{kj}^* V_{kl} V_{il}^*$ ($i \neq k$, $j \neq l$), are
equal up to a sign. In particular, one has
\begin{eqnarray}
T_1 & \equiv & {\mathrm Im}\;V_{ud} V_{cd}^* V_{cb} V_{ub}^*+{\mathrm Im}\;
V_{us} V_{cs}^* V_{cb} V_{ub}^* = 0\,, \nonumber  \\
T_2 & \equiv & {\mathrm Im}\;V_{ud} V_{td}^* V_{tb} V_{ub}^*+{\mathrm Im}\;
V_{us} V_{ts}^* V_{tb} V_{ub}^* = 0\,, \nonumber  \\
T_3 & \equiv & {\mathrm Im}\;V_{cd} V_{td}^* V_{tb} V_{cb}^*+{\mathrm Im}\;
V_{cs} V_{ts}^* V_{tb} V_{cb}^* = 0\,. 
\label{ec:1}
\end{eqnarray}
In the SM one may have CP violation in the limit $m_{u,d}=0$, but degeneracy of
two quarks of the same charge does imply CP invariance. Hence in the chiral
limit $m_{u,d,s}=0$, with $d$ and $s$ quark masses degenerate, CP
violation can only originate in physics beyond the SM. These CP properties of
the SM are summarized through a necessary and sufficient condition for CP
invariance, expressed in terms of a weak quark basis invariant
\cite{papiro4,papiro4b}
\begin{eqnarray}
I & \equiv & \det\; [M_u M_u^\dagger,M_d M_d^\dagger]
 =  \frac{1}{3}{\mathrm tr\;}[M_u M_u^\dagger,M_d M_d^\dagger]^3 \nonumber \\
& = & -2 i (m_t^2-m_c^2)(m_t^2-m_u^2) (m_c^2-m_u^2)
(m_b^2-m_s^2) \nonumber \\
& & \times (m_b^2-m_d^2)(m_s^2-m_d^2)\,{\mathrm Im}\;
V_{ud} V_{cd}^* V_{cs} V_{us}^{*}=0\,,
\label{ec:2}
\end{eqnarray}
where $M_{u,d}$ are the up and down quark mass matrices and $m_i$ the mass of
the quark $i$. Although for three generations the two invariants of Eq.
(\ref{ec:2}) are proportional to each other, ${\mathrm tr}\; [M_u M_u^\dagger,
M_d M_d^\dagger]^3 = 0$ has the advantage of being a nontrivial necessary
condition for CP invariance in the SM, for
an arbitrary number of generations \cite{papiro4b}.

Within the three generation
SM, CP violation in the B system
is not suppressed by the factor $(m_s-m_d)$ due
to the fact that one is able to
distinguish ${\mathrm B}_d$ from ${\mathrm B}_s$
in the initial state and kaons from pions in the final state. This flavour
identification is crucial in order to detect CP violation effects arising from
gauge interactions \cite{papiro5b,papiro5c,papiro5d,papiro5}. At high energy
colliders, the natural asymptotic states are no longer
 hadronic states, but quark
jets \cite{papironn1}. Now, at high energies, it will be very hard, if not
impossible, to identify the flavour of light quark jets ($d$, $s$, $u$). In this
limit, CP violation can only originate in physics beyond the SM. Indeed many
simple SM extensions incorporate new sources of CP violation
which may be observable at future colliders in the chiral limit
\cite{papiro12}.
In this paper we concentrate on the case of
extra quarks, with special emphasis on vector-like quarks, {\em i. e.} quarks
whose left-handed and right-handed components are in the same type of multiplet.
Vector-like quarks naturally arise in various extensions of the SM, for example
in grand-unified theories based on ${\mathrm E}_6$, as well as in other
superstring inspired extensions of the SM. In most cases, these vector-like
quarks are isosinglets and their mass could be of the order of the electroweak
scale \cite{papiro2}. A new sequential quark family is also allowed
\cite{papiro7b}, although
precision electroweak data put an upper limit on their square mass difference
\cite{papiro2}. We will show in this paper that with the adition of an extra
quark, either sequential or vector-like, there is CP violation even in the limit
where $m_{u,d,s,c}=0$. Therefore, in these simple extensions of the SM, CP
violation effects can be seen even if only the flavour of heavy quarks ($b$, $t$,
$b'$) is identified. In both cases, for an isosinglet quark and for a
sequential extra family, CP violation mediated by gauge bosons is parametrized
by a $4 \times 4$ unitary matrix $V$ defined up to quark mass eigenstate phase
redefinitions. For a new down (up) vector-like quark the charged couplings are
described by the CKM matrix $V_{\mathrm CKM}$, the first 3 rows (columns) of
$V$. But these 3 rows (columns) $V_{\mathrm CKM}$
completely fix $V$. The neutral
couplings are a function of the $V_{\mathrm CKM}$ matrix elements and are not
independent. In the case of an isosinglet quark, the neutral couplings are no
longer diagonal, {\em i. e.} there are flavour changing neutral currents
(FCNC). The $3 \times 3$ block of the CKM matrix connecting standard quarks is
no longer unitary either,
but deviations from unitarity are naturally suppressed by
powers of $m/M$, where $m$ is a standard quark mass and $M$ denotes the mass of
the isosinglet quark. The strength of FCNC among standard quarks is also
suppressed by powers of $m/M$, since FCNC are proportional to deviations from
$3 \times 3$ unitarity in $V_{\mathrm CKM}$.
For a new sequential family $V_{\mathrm CKM}=V$ and the neutral couplings are
diagonal and real. It turns out that in both cases there are three CP violating
phases in $V_{\mathrm CKM}$ \cite{papiro10}. 

In our analysis we will adopt the following strategy: First, we identify a set
of weak-basis invariants which completely specify the properties of the model
considered in the sense that if any one of the invariants is nonzero there is
CP violation, while if all the invariants of the set vanish there is CP
invariance. These weak-basis invariants are physically meaningful quantities,
and they are the analog of the invariant in Eq. (\ref{ec:2}) for the class of
models we are considering. They can be expressed in terms of quark masses and
the imaginary parts of various invariant products of CKM matrix elements.
We then study the chiral
limit $m_{u,d,s}=0$, starting with the simpler case
$m_{u,d,s,c}=0$. Both are especially interesting since
in these cases the three  generation SM conserves CP, thus implying that in the
chiral limit CP violation arises exclusively from physics beyond the SM. 
The chiral limit is not only physically natural for studying CP violation
beyond the SM but phenomenologically relevant at high energy, where the light
fermion masses are negligible. The
above mentioned weak-basis invariants are especially useful in the analysis of
the chiral limit, allowing one to readily identify which 
${\mathrm Im}\;V_{ij} V_{kj}^* V_{kl} V_{il}^*$ continue being physically
meaningful and nonvanishing when taking these degenerate mass limits.
In a $4 \times 4$ unitary matrix 9 of these imaginary products
are independent, and all of them can be made to vanish if CP is conserved. In
the chiral limit $m_{u,d,s}=0$, there are two CP violating phases and three
independent imaginary products physically relevant.
If $m_c$ is also neglected compared to
$m_t$, $m_{u,d,s,c}=0$, there is one CP violating phase left and one
independent imaginary product physically significant. These
imaginary products 
\begin{eqnarray}
B_1 & \equiv & {\mathrm Im}\; V_{cb} V_{4b}^* V_{4b'} V_{cb'}^* = T_1 - T_3
\,,\nonumber \\
B_2 & \equiv & {\mathrm Im}\; V_{tb} V_{4b}^* V_{4b'} V_{tb'}^* = T_2 + T_3
\,,\nonumber \\
B_3 & \equiv & {\mathrm Im}\; V_{cb} V_{tb}^* V_{tb'} V_{cb'}^* = T_3 \,,
\label{ec:n6}
\end{eqnarray}
which survive in the chiral limit and which involve mixings
in the heavy quark sector ($t$, $b$, $c$ and the new quark(s) ),
can also be expressed
in terms of the rephasing invariants $T_i$ defined in Eqs. (\ref{ec:1}). 
(We will use for the subindex of the fourth row `4' and `$t'$' when referring to
the vector-like and sequential cases, respectively. When referring to both we
will also use `4'.) 
These
invariants $T_i$ only involve mixings among standard quarks and they vanish in
the SM. Thus, the effects of physics beyond the SM may be seen measuring $B_i$
at high
energy in processes involving new quarks or at low energy through the
nonvanishing of $T_i$. We shall show that the present bounds on $|B_i|$ are
$10^{-2}$ and $10^{-4}$ for a fourth family and a new vector-like quark,
respectively. 
In our analysis, we will only take into account CP violating effects arising
from gauge interactions (in some specific models with isosinglet quarks the
Higgs sector is more involved than in the SM, leading to new CP violating
contributions from scalar interactions) and we will neglect the
Higgs contributions to CP violation.

It may be worth to emphasize that
the invariant formulation of CP violation requires an educated
use of symbolic programs \cite{papiroa2}. However to go beyond the simplest
cases is difficult for as explained in the Appendix the number of invariants
needed to get a complete set grows very rapidly with the number of phases.
We study the simplest cases of a new vector-like quark or an extra sequential
family, deriving limits for observables
involving known particles as final states.
If there exist more vector-like or sequential
quarks, larger CP violating effects than the ones
studied here are possible but in observables involving several of these new
quarks.

This paper is organized as follows. In Section 2 we set
our notation and for the case of an extra down (up) quark isosinglet we propose
a complete set of weak-basis invariant conditions which are necessary and
sufficient for CP invariance. We also give the explicit expressions of the
invariants in terms of quark masses and imaginary parts of invariant products.
The proof that these invariant conditions form a complete set is
given in the Appendix. The corresponding set of invariants for the case of a
sequential family was discussed in Ref. \cite{papiro8}. In Section 3 we use
these invariants to study the simplest case of two
degenerate (massless) up and down quark masses, $m_{u,d,s,c}=0$, and the
chiral limit, $m_{u,d,s}=0$. In Section 4 we discuss the corresponding
geometrical description of CP violation with triangles and quadrangles and
the bounds on the
CP violating effects of these new fermions commenting on
the prospects to measure  
them (at large colliders). Section 5 is devoted to our conclusions.

\section{Characterization of CP violation for extra quarks}
Let us consider the SM with $N$ standard families plus $n_{d}$ down
quark isosinglets (the case of $n_u$ up quark isosinglets is similar).
In the weak eigenstate basis the gauge couplings to quarks and the mass terms
are
\begin{eqnarray}
{\mathcal L}_{\mathrm gauge}& = & -\frac{g}{\sqrt 2}
\left [  \left( \bar u_{L}^{(d)} \gamma^\mu d_{L}^{(d)}
\right) W_\mu^\dagger + {\mathrm h.c.} \right ] \nonumber \\
& & -\frac{g}{2c_{\mathrm W}} \left (  \bar u_{L}^{(d)}
\gamma^\mu u_{L}^{(d)}-
\bar d_{L}^{(d)} \gamma^\mu d_{L}^{(d)} 
 - 2 s_{\mathrm W}^2 J_{\mathrm EM}^\mu \right )
Z_{\mu} -e J_{\mathrm EM}^\mu A_\mu \,, \label{ec:4a} \\
{\mathcal L}_{\mathrm mass} & = & -\left( \bar u_L^{(d)} M_u u_R^{(s)}
+ \bar d_L^{(d)} M_d d_R^{(s)} + \bar d_L^{(s)} m_d d_R^{(s)} \right) +
{\mathrm h.c.} \,,
\label{ec:4b}
\end{eqnarray}
where $u_L^{(d)}$, 
$d_L^{(d)}$ are $N$ ${\mathrm SU}(2)_L$ doublets, $d_L^{(s)}$ are $n_d$
${\mathrm SU}(2)_L$ singlets and $u_R^{(s)}$ and $d_R^{(s)}$ are $N$ and $N+n_d$
${\mathrm SU}(2)_L$ singlets, respectively, and
$J_{\mathrm EM}^\mu = \frac{2}{3} \bar u \gamma^\mu u-\frac{1}{3} \bar d
\gamma^\mu d$. Hence, the up and down quark mass matrices are
\begin{equation}
{\mathcal M}_u=M_u ~,~~
{\mathcal M}_d = \left( \begin{array}{c}
M_d  \\ m_d  \end{array} \right)\,,
\label{ec:5}
\end{equation}
with $M_u$, $M_d$ and $m_d$ submatrices of dimension $N \times N$, $N \times
(N+n_d)$ and $n_d \times (N+n_d)$, respectively. The weak
quark basis can be transformed by unitary matrices without changing
the physics. Under these unitary transformations
\begin{equation}
q_{L}^{(d)} \rightarrow U_L  q_{L}^{(d)}  ~,~~ 
d_{L}^{(s)} \rightarrow U_L^d d_{L}^{(s)} ~,~~ 
q_{R}^{(s)} \rightarrow U_R^q q_{R}^{(s)} \,,
\label{ec:6}
\end{equation}
with $q=u,d$, the mass matrices transform as
\begin{equation}
M_q \rightarrow U_L M_q U_R^{q\dagger} ~,~~
m_d \rightarrow U_L^d m_d U_R^{d\dagger} \,,
\label{ec:6bis}
\end{equation}
whereas the gauge couplings remain unchanged. Then
the mass matrices are defined up
to the unitary transformations in Eq. (\ref{ec:6bis}). 

A set of physical parameters can be defined using the mass eigenstate basis.
We assume without loss of generality ${\mathcal M}_u = {\mathcal D}_u$
diagonal and ${\mathcal M}_d
= V {\mathcal D}_d  V^\dagger$, with $V$ unitary and ${\mathcal D}_d$
diagonal (remember that we can always assume ${\mathcal M}_u$ and
${\mathcal M}_d$ hermitian with nonnegative eigenvalues by choosing $U_R^{q}$
appropriately). We define the $N \times (N+n_d)$ matrix  $V_{\mathrm CKM}$ as
the first $N$ rows of the $(N+n_d) \times (N+n_d)$ unitary matrix $V$, and the
$(N+n_d) \times (N+n_d)$ matrix $X \equiv V_{\mathrm CKM}^\dagger
V_{\mathrm CKM}$.
Then the Lagrangian in Eqs. (\ref{ec:4a},\ref{ec:4b}) reads in the
quark mass eigenstate basis (where no superscripts are needed)
\begin{eqnarray}
{\mathcal L}_{\mathrm gauge} & = & - \frac{g}{\sqrt 2} \left(
\bar u_L \gamma^\mu V_{\mathrm CKM} d_L W_\mu^\dagger+{\mathrm h.c} \right)
\nonumber \\
& & -\frac{g}{2c_{\mathrm W}} \left( \bar u_L \gamma^\mu u_L - \bar d_L
\gamma^\mu X d_L 
- 2 s_{\mathrm W}^2 J_{\mathrm EM}^\mu \right) Z_\mu
-e J_{\mathrm EM}^\mu A_\mu \,,
\label{ec:13a} \\
{\mathcal L}_{\mathrm mass} & = & - \left( \bar u_L {\mathcal D}_u u_R
+ \bar d_L {\mathcal D}_d d_R \right)  +{\mathrm h.c.} \,.
\label{ec:13b}
\end{eqnarray}
In this basis one can make the counting of CP violating phases. This is
equal to the number of phases in $V_{\mathrm CKM}$ minus the number of
independent phase field redefinitions \cite{papiro6b},
\begin{eqnarray}
n_{\mathrm CP} & = & N(N+n_d)-\frac{N(N-1)}{2}-(2N+n_d-1) \nonumber \\
& = & (N-1)n_d+\frac{1}{2}(N-1)(N-2) \,.
\label{ec:14}
\end{eqnarray}
CP is conserved if $V_{\mathrm CKM}$ can be made real. This is the case if all
$n_{\mathrm CP}$ phases vanish. In the SM, $N=3$, $n_d=0$, there is only 1 CP
violating phase. For one extra quark, $N=3$, $n_d=1$, there are already 3 CP
violating phases, the same as for an extra family, $N=4$, $n_d=0$. The number
of physical parameters, and in particular of CP violating phases, grows
rapidly with the addition of more quarks. We will stick to these cases which
incorporate many of the new features of the addition of new quark fields. For
an extra down quark isosinglet, $V_{\mathrm CKM}$ consists of
the first 3 rows of a $4 \times 4$
unitary matrix which can be parametrized as
(we explicit it for later use) \cite{papiro10}
\begin{equation}
V=\left( 
\begin{scriptsize}
\begin{array}{cccc}
c_1 & s_1 c_3 & s_1 s_3 c_5 & s_1 s_3 s_5 \\
-s_1 c_2 & c_1 c_2 c_3 + s_2 s_3 c_6 e^{i \delta_1} &
c_1 c_2 s_3 c_5 - s_2 c_3 c_5 c_6 e^{i \delta_1} &
c_1 c_2 s_3 s_5 - s_2 c_3 s_5 c_6 e^{i \delta_1} \\
& & +s_2 s_5 s_6 e^{i(\delta_1 + \delta_3)} &
-s_2 c_5 s_6 e^{i(\delta_1+\delta_3)} \\
-s_1 s_2 c_4 & c_1 s_2 c_3 c_4 - c_2 s_3 c_4 c_6 e^{i \delta_1} &
c_1 s_2 s_3 c_4 c_5 + c_2 c_3 c_4 c_5 c_6 e^{i \delta_1} &
c_1 s_2 s_3 c_4 s_5 + c_2 c_3 c_4 s_5 c_6 e^{i \delta_1} \\
& -s_3 s_4 s_6 e^{\i \delta_2} & -c_2 c_4 s_5 s_6 e^{i(\delta_1+\delta_3)} &
+c_2 c_4 c_5 s_6 e^{i(\delta_1+\delta_3)} \\
& & +c_3 s_4 c_5 s_6 e^{i \delta_2} & +c_3 s_4 s_5 s_6 e^{i\delta_2} \\
& & +s_4 s_5 c_6 e^{i(\delta_2+\delta_3)} & -s_4 c_5 c_6
e^{i(\delta_2+\delta_3)} \\
-s_1 s_2 s_4 & c_1 s_2 c_3 s_4 - c_2 s_3 s_4 c_6 e^{i\delta_1} &
c_1 s_2 s_3 s_4 c_5 + c_2 c_3 s_4 c_5 c_6 e^{i\delta_1} &
c_1 s_2 s_3 s_4 s_5 + c_2 c_3 s_4 s_5 c_6 e^{i\delta_1} \\
& +s_3 c_4 s_6 e^{i\delta_2} & -c_2 s_4 s_5 s_6 e^{i(\delta_1+\delta_3)} &
+c_2 s_4 c_5 s_6 e^{i(\delta_1+\delta_3)} \\
& & -c_3 c_4 c_5 s_6 e^{i\delta_2} & -c_3 c_4 s_5 s_6 e^{i\delta_2} \\
& & -c_4 s_5 c_6 e^{i(\delta_2+\delta_3)} & +c_4 c_5 c_6
e^{i(\delta_2+\delta_3)}
\end{array}
\end{scriptsize}
\right)\,.
\label{ec:14bis}
\end{equation}
Note that the first 3 rows completely fix $V$. On the other hand, $V$ is the CKM
matrix for 4 families. In the limit where the new quark does not mix with the
standard quarks ({\em i. e.} $s_4=s_5=s_6=0$), the $3 \times 3$ block of $V$
becomes just the standard CKM matrix with only one CP violating phase
$\delta_1$. The CP properties of the model with one isosinglet quark can be
most conveniently studied by using weak-basis invariants. In the Appendix we
present the general treatment and provide the proof that the vanishing of the
following set of
invariants is necessary and sufficient to have CP invariance:
\begin{eqnarray}
I_1 & = & {\mathrm Im~tr}\;H_u H_d h_d h_d^\dagger \,, \nonumber \\
I_2 & = & {\mathrm Im~tr}\;H_u^2 H_d h_d h_d^\dagger \,, \nonumber \\
I_3 & = & {\mathrm Im~tr}\;(H_u^3 H_d h_d h_d^\dagger
- H_u^2 H_d H_u h_d h_d^\dagger) \,, \nonumber \\
I_4 & = & {\mathrm Im~tr}\;H_u H_d^2 h_d h_d^\dagger \,, \nonumber \\
I_5 & = & {\mathrm Im~tr}\;H_u^2 H_d^2 h_d h_d^\dagger \,, \nonumber \\
I_6 & = & {\mathrm Im~tr}\;(H_u^3 H_d^2 h_d h_d^\dagger
- H_u^2 H_d^2 H_u h_d h_d^\dagger) \,, \nonumber \\
I_7 & = & {\mathrm Im~tr}\;H_u^2 H_d H_u H_d^2 \,,
\label{ec:16}
\end{eqnarray}
with $H_u=M_u M_u^\dagger$, $H_d=M_d M_d^\dagger$, $h_d=M_d m_d^\dagger$ (see
Eq. (\ref{ec:5}) ). These
invariants, which obviously do not depend on the choice of weak quark basis
(see Eq. (\ref{ec:6bis}) ),
can be written in the quark mass eigenstate basis (in the equations below Greek
indices run from 1 to 4, Latin
indices run from 1 to 3 and a sum over all indices is implicit; $m_i$ is the
mass of the up quark $i$ and $n_\alpha$ the mass of the down quark $\alpha$)
\begin{eqnarray}
I_1 & = & m_i^2 n_\alpha^2 n_\beta^4 \;{\mathrm Im}\;V_{i \alpha} X_{\alpha
\beta} V_{i \beta}^*\,, \nonumber \\
I_2 & = & m_i^4 n_\alpha^2 n_\beta^4 \;{\mathrm Im}\;V_{i \alpha} X_{\alpha
\beta} V_{i \beta}^*\,, \nonumber \\
I_3 & = & m_i^6 n_\alpha^2 n_\beta^4 \;{\mathrm Im}\; V_{i \alpha}
X_{\alpha \beta} V_{i \beta}^* \nonumber \\
& & - m_i^4 m_j^2 n_\alpha^2 n_\beta^4 \; {\mathrm Im}\;V_{i \alpha} V_{j
\alpha}^* V_{j \beta} V_{i \beta}^* \nonumber \\
& & + m_i^4 m_j^2 n_\alpha^2 n_\beta^2 n_\rho^2 \; {\mathrm Im}\; V_{i \rho}
V_{j \rho}^* V_{j \alpha} X_{\alpha \beta} V_{i \beta}^*\,, \nonumber \\
I_4 & = & m_i^2 n_\alpha^2 n_\beta^4 n_\rho^2 \;{\mathrm Im}\; V_{i \rho}
X_{\rho \alpha} X_{\alpha \beta} V_{i \beta}^*\,, \nonumber \\
I_5 & = & m_i^4 n_\alpha^2 n_\beta^4 n_\rho^2 \;{\mathrm Im}\; V_{i \rho}
X_{\rho \alpha} X_{\alpha \beta} V_{i \beta}^*\,, \nonumber \\
I_6 & = & m_i^6 n_\alpha^2 n_\beta^4 n_\rho^2 \;{\mathrm Im}\; V_{i \rho}
X_{\rho \alpha} X_{\alpha \beta} V_{i \beta}^* \nonumber \\
& & - m_i^4 m_j^2 n_\alpha^2 n_\beta^2 n_\rho^4 \;{\mathrm Im}\; V_{i \alpha}
X_{\alpha \beta} V_{j \beta}^* V_{j \rho} V_{i \rho}^*\,, \nonumber \\
I_7 & = & m_i^4 m_j^2 n_\alpha^2 n_\beta^2 n_\rho^2 \;{\mathrm Im}\; V_{i \rho}
V_{j \rho}^* V_{j \alpha} X_{\alpha \beta} V_{i \beta}^*\,.
\label{ec:17}
\end{eqnarray}
Notice that $X_{\alpha \beta}=V_{i\alpha}^* V_{i\beta}$, which implies
$X=X^\dagger$ and
$X_{\alpha \beta} X_{\beta \gamma} = X_{\alpha \gamma}$ (note however that
in Eqs. (\ref{ec:17}) the sums include also mass factors). The imaginary parts
involve invariant quartets and invariant sextets, which can be reduced also to
products of moduli squared of $V_{ij}$ elements times imaginary parts of
quartets. In the chiral limits there only appear imaginary parts of invariant
quartets. 
For the case of
four sequential families, the corresponding set of necessary and sufficient
conditions for CP invariance consists of eight invariants which have been given
in Ref. \cite{papiro8}. In both cases these invariant conditions completely
characterize the CP properties of the model. If any of the invariants is
nonvanishing, there is CP violation and the vanishing of the invariants
implies CP invariance. The description of the CP properties of a model through
invariants is especially useful when considering limiting cases where some of
the quark masses can be considered as degenerate (massless).
The invariant approach
clearly identifies which ones of the 
${\mathrm Im}\;V_{ij} V_{kj}^* V_{kl} V_{il}^*$ can be nonvanishing in the
various limiting cases one considers. The difficult task is, of course, finding
a complete set of invariants, but once this is accomplished, the invariant
approach is a very convenient method to describe CP violation. In the next
Section we will illustrate the usefulness of these invariant conditions, by
considering some appropriate chiral limits.

\section{CP violation in the chiral limit}
In the chiral limit, $m_{u,d,s}=0$, CP is conserved within the SM. Hence all CP
violating effects are due to new physics. Sizeable CP violation at high energy
is expected to have its origin beyond the SM. Let us discuss in turn the
simplest limit of $m_t \gg m_c \sim 0$, $m_{u,d,s}=0$ and the chiral limit
$m_{u,d,s}=0$. We will consider the cases of an extra isosinglet quark and of a
fourth sequential family.

\subsection{$m_{u,d,s,c}=0$ limit}
In this limit there is only one CP violating phase for an extra down quark
isosinglet $b'$ or for a fourth family $b'$, $t'$.
The best way to study this limit
is to substitute $m_{u,d,s,c}=0$ in the complete set of invariants
characterizing CP. The 7 invariants in Eqs. (\ref{ec:16}) reduce to
\begin{eqnarray}
I_1 & = & m_t^2 m_{b'}^2 m_b^2 (m_{b'}^2-m_b^2)\, {\mathrm Im}\; V_{tb} X_{bb'}
V_{tb'}^* \,, \nonumber \\
I_2 & = & m_t^2 I_1 \,, \nonumber \\
I_3 & = & m_t^4 I_1 \,, \nonumber \\
I_4 & = & (m_{b'}^2 X_{b'b'} + m_b^2 X_{bb}) I_1 \,, \nonumber \\
I_5 & = & m_t^2 I_4 \,, \nonumber \\
I_6 & = & m_t^4 I_4 \,, \nonumber \\
I_7 & = & 0 \,.
\label{ec:18}
\end{eqnarray}
It is clear that CP is conserved if and only if ${\mathrm Im}\; V_{tb} X_{bb'}
V_{tb'}^*=0$. Obviously, we have made the assumption that $b$, $b'$ are
nondegenerate, with $m_{b'} > m_b$. Hence all CP violating
effects are proportional to this imaginary product which gives the size of CP
violation. Similarly, for 4 families the corresponding 8 invariants 
\cite{papiro8} reduce to (we use a prime to distinguish them)
\begin{eqnarray}
I'_1 & = & -m_{t'}^2 m_t^2 (m_{t'}^2-m_t^2) m_{b`}^2 m_b^2 (m_{b'}^2-m_b^2)\,
{\mathrm Im}\; V_{tb} V_{t'b}^* V_{t'b'} V_{tb'}^* \,, \nonumber \\
I'_2 & = & (m_{t'}^2+m_t^2) I'_1 \,, \nonumber \\
I'_3 & = & (m_{t'}^4+m_t^4) I'_1 \,, \nonumber \\
I'_4 & = & (m_{t'}^6+m_t^6) I'_1 \,, \nonumber \\
I'_5 & = & (m_{b'}^2+m_b^2) I'_1 \,, \nonumber \\
I'_6 & = & (m_{b'}^2+m_b^2) I'_2 \,, \nonumber \\
I'_7 & = & (m_{b'}^4+m_b^4) I'_1 \,, \nonumber \\
I'_8 & = & (m_{b'}^6+m_b^6) I'_1 \,.
\label{ec:19}
\end{eqnarray}
In this case CP conservation reduces to requiring ${\mathrm Im}\; V_{tb}
V_{t'b}^* V_{t'b'} V_{tb'}^*=0$, with the implicit assumption
$m_{b'}>m_b$ and $m_{t'}>m_t$. Not only the
same comments as for an extra isosinglet apply but the observable
for SM final states is the same. Thus using unitarity
\begin{equation}
B_2 = {\mathrm Im}\; V_{tb} V_{4b}^* V_{4b'} V_{tb'}^* 
= - {\mathrm Im}\; V_{tb} X_{bb'} V_{tb'}^*
= - {\mathrm Im}\; V_{tb} V_{ib}^* V_{ib'} V_{tb'}^* \,.
\label{ec:20}
\end{equation}
It is clear that $B_2$ measures the strength of CP violation in high energy
processes involving the new quark. On the other hand due to the unitarity
constraints, $B_2$ is actually related to the invariants $T_i$ defined in Eqs.
(\ref{ec:1}), which only depend on standard quark mixings. Indeed, one obtains
\begin{eqnarray}
B_2 = -{\mathrm Im}\; V_{tb} V_{ib}^* V_{ib'} V_{tb'}^* & =  &-{\mathrm Im}\; 
V_{tb} V_{ub}^* V_{ub'} V_{tb'}^*-{\mathrm Im}\; V_{tb} V_{cb}^* V_{cb'} 
V_{tb'}^* \nonumber \\
& = & {\mathrm Im}\; V_{tb} V_{ub}^* V_{ud} V_{td}^* +
{\mathrm Im}\; V_{tb} V_{ub}^* V_{us} V_{ts}^* \nonumber \\
& & +{\mathrm Im}\; V_{tb} V_{cb}^* V_{cd} V_{td}^* +
{\mathrm Im}\; V_{tb} V_{cb}^* V_{cs} V_{ts}^* = T_2 + T_3 \,,
\label{ec:21}
\end{eqnarray}
with $T_{2,3}=0$ in the three generation SM as emphasized in the Introduction.

All this can also be proven using the CKM matrix, although the physics is less
transparent. If $m_{u,c}=0$, $m_{d,s}=0$, the general $4 \times 4$ unitary
matrix in Eq. 
(\ref{ec:14bis}) (up to quark field phase redefinitions) can be written
\begin{equation}
V=\left( \begin{array}{cccc}
c_1 & s_1 c_3 & s_1 s_3 & 0 \\
-s_1 c_2 & c_1 c_2 c_3 + s_2 s_3 c_6 e^{i\delta_1} &
c_1 c_2 s_3 - s_2 c_3 c_6 e^{i\delta_1} & -s_2 s_6 e^{i\delta_1} \\
-s_1 s_2 & c_1 s_2 c_3-c_2 s_3 c_6 e^{i\delta_1} & 
c_1 s_2 s_3 + c_2 c_3 c_6 e^{i\delta_1} & c_2 s_6 e^{i\delta_1} \\
0 & s_3 s_6 & -c_3 s_6 & c_6 
\end{array}
\right ) \,,
\label{ec:22}
\end{equation}
where we have used the freedom to make unitary transformations in the ($u$,$c$)
and ($d$,$s$) spaces. This freedom results of course from the fact that 
($u$,$c$) and ($d$,$s$) are degenerate in the limit we are considering. As
expected, in this limit there is only one CP violating phase and the strength
of CP violation is given by
\begin{equation}
B_2 = {\mathrm Im}\;V_{tb} V_{4b}^* V_{4b'} V_{tb'}^* = c_1 c_2 s_2 c_3 s_3 c_6
s_6^2 \sin \delta_1 \,.
\label{ec:23}
\end{equation}
Unitarity allows to recover Eqs. (\ref{ec:20},\ref{ec:21}). In
this particular parametrization $B_2$ is in fact equal to $T_3$ for $T_{1,2}=0$
as defined in Eqs. (\ref{ec:1}). 


\subsection{The chiral limit, $m_{u,d,s}=0$}
In this limit there is no CP violation in the SM, but for one extra quark
isosinglet or a fourth sequential family there are two new CP violating phases
which remain physical. In the case of the model with an extra down quark
isosinglet, CP conservation is equivalent to the vanishing of $I_{1-3}$ in Eqs.
(\ref{ec:16}), because in this limit
\begin{eqnarray}
I_1 & = & m_t^2  I_t + m_c^2 I_c \,, \nonumber \\
I_2 & = & m_t^4 I_t + m_c^4 I_c \,, \nonumber \\
I_3 & = & m_t^6 I_t + m_c^6 I_c + m_t^2 m_c^2 (m_t^2-m_c^2) m_{b'}^2
m_b^2 (m_{b'}^2-m_b^2)\, {\mathrm Im}\; V_{cb} V_{tb}^* V_{tb'} V_{cb'}^*
+ I_7 \,, \nonumber \\
I_4 & = & (m_{b'}^2 X_{b'b'}+m_b^2 X_{bb})\, I_1 \,, \nonumber \\
I_5 & = & (m_{b'}^2 X_{b'b'}+m_b^2 X_{bb})\, I_2 \,, \nonumber \\
I_6 & = & (m_{b'}^2 X_{b'b'}+m_b^2 X_{bb})\, (m_t^6 I_t + m_c^6 I_c)
\nonumber \\
& & - m_t^2 m_c^2 (m_t^2-m_c^2) m_{b'}^2 m_b^2  [ m_{b'}^2 m_b^2
(X_{b'b'}-X_{bb})  {\mathrm Im}\; V_{cb} V_{tb}^* V_{tb'} V_{cb'}^* \nonumber \\
& & + (m_{b'}^4 |V_{cb'}|^2-m_b^4 |V_{cb}|^2) {\mathrm Im}\;
V_{tb} X_{bb'} V_{tb'}^* \nonumber \\
& & - (m_{b'}^4 |V_{tb'}|^2 -m_b^4 |V_{tb}|^2) {\mathrm Im}\; 
V_{cb} X_{bb'} V_{cb'}^*  ] \,, \nonumber \\
I_7 & = & - m_t^2 m_c^2 (m_t^2-m_c^2) m_{b'}^2 m_b^2 [ (m_{b'}^2 X_{b'b'} -m_b^2
X_{bb}) {\mathrm Im}\; V_{cb} V_{tb}^* V_{tb'} V_{cb'}^* \nonumber \\
& & + (m_{b'}^2 |V_{cb'}|^2-m_b^2 |V_{cb}|^2) {\mathrm Im}\;
V_{tb} X_{bb'} V_{tb'}^* \nonumber \\
& & - (m_{b'}^2 |V_{tb'}|^2-m_b^2 |V_{tb}|^2) {\mathrm Im}\;
V_{cb} X_{bb'} V_{cb'}^* ]\,,
\label{ec:24}
\end{eqnarray}
with
\begin{eqnarray}
I_c & = & m_c^2 m_{b'}^2 m_b^2 (m_{b'}^2-m_b^2)\, {\mathrm Im}\; V_{cb} X_{bb'}
V_{cb'}^* \,, \nonumber \\
I_t & = & m_t^2 m_{b'}^2 m_b^2 (m_{b'}^2-m_b^2)\, {\mathrm Im}\; V_{tb} X_{bb'}
V_{tb'}^* \,.
\label{ec:24bis}
\end{eqnarray}
There are only three independent imaginary products, ${\mathrm Im}\; V_{tb} X_{bb'}
V_{tb'}^*$, ${\mathrm Im}\; V_{cb} X_{bb'} V_{cb'}^*$ and ${\mathrm Im}\;
V_{cb} V_{tb}^* V_{tb'} V_{cb'}^*$, entering in $I_{1-7}$. Their vanishing
guarantees CP conservation. The first one is the only one which survives for
$m_c=m_u$ as proven in the previous Subsection. Analogously, in the case of a
fourth family CP invariance is equivalent to the vanishing of $I'_{1-3}$ in
Ref. \cite{papiro8}. In this case the complete set of 8 invariants reduces to 
\begin{eqnarray}
I'_1 & = & I'_{ct}+I'_{ct'}+I'_{tt'} \,, \nonumber \\
I'_2 & = & (m_t^2+m_c^2) I'_{ct} + (m_{t'}^2+m_c^2) I'_{ct'} +
(m_{t'}^2+m_t^2) I'_{tt'} \,,\nonumber \\
I'_3 & = & (m_t^4+m_c^4) I'_{ct} + (m_{t'}^4+m_c^4) I'_{ct'} +
(m_{t'}^4+m_t^4) I'_{tt'} \,,\nonumber \\
I'_4 & = & (m_t^6+m_c^6) I'_{ct} + (m_{t'}^6+m_c^6) I'_{ct'} +
(m_{t'}^6+m_t^6) I'_{tt'} \,,\nonumber \\
& & - m_{t'}^2 m_t^2 m_c^2 (m_t^2-m_c^2) (m_{t'}^2-m_c^2) (m_{t'}^2-m_t^2)
m_{b'}^2 m_b^2 \nonumber \\
& & \times [ (|V_{t'b'}|^2 m_{b'}^2-|V_{t'b}|^2 m_b^2)\, {\mathrm Im}\; V_{cb}
V_{tb}^* V_{tb'} V_{cb'}^* \nonumber \\
& & - (|V_{tb'}|^2 m_{b'}^2-|V_{tb}|^2 m_b^2)\, {\mathrm Im}\; V_{cb} V_{t'b}^*
V_{t'b'} V_{cb'}^* \nonumber \\
& & + (|V_{cb'}|^2 m_{b'}^2-|V_{cb}|^2 m_b^2)\, {\mathrm Im}\; V_{tb} V_{t'b}^*
V_{t'b'} V_{tb'}^* ] \,,\nonumber \\
I'_5 & = & (m_{b'}^2+m_b^2) I'_1 \,, \nonumber \\
I'_6 & = & (m_{b'}^2+m_b^2) I'_2 \,, \nonumber \\
I'_7 & = & (m_{b'}^4+m_b^4) I'_1 \,, \nonumber \\
I'_8 & = & (m_{b'}^6+m_b^6) I'_1 \,,
\label{ec:25}
\end{eqnarray}
with
\begin{eqnarray}
I'_{ct} & = & - m_{t}^2 m_{c}^2 (m_{t}^2-m_{c}^2) m_{b'}^2 m_b^2
(m_{b'}^2-m_b^2) {\mathrm Im}\; V_{cb} V_{tb}^* V_{tb'} V_{cb'}^* \,, 
\nonumber \\
I'_{ct'} & = & - m_{t'}^2 m_{c}^2 (m_{t'}^2-m_{c}^2) m_{b'}^2 m_b^2
(m_{b'}^2-m_b^2) {\mathrm Im}\; V_{cb} V_{t'b}^* V_{t'b'} V_{cb'}^* \,,
\nonumber \\
I'_{tt'} & = & - m_{t'}^2 m_{t}^2 (m_{t'}^2-m_{t}^2) m_{b'}^2 m_b^2
(m_{b'}^2-m_b^2) {\mathrm Im}\; V_{tb} V_{t'b}^* V_{t'b'} V_{tb'}^* \,.
\label{ec:25bis}
\end{eqnarray}

For four families there are also three independent
imaginary products, ${\mathrm Im}\;
V_{tb} V_{t'b}^* V_{t'b'} V_{tb'}^*$, ${\mathrm Im}\; V_{cb} V_{t'b}^* V_{t'b'}
V_{cb'}^*$ and ${\mathrm Im}\; V_{cb} V_{tb}^* V_{tb'} V_{cb'}^*$, entering in
$I'_{1-8}$. Similarly to the case of an extra isosinglet their vanishing
guarantees CP conservation. 
There is an interesting connection between these rephasing invariants $B_i$,
which can see CP violation through high energy processes involving the
new quark, and $T_i$, which only involve mixings among SM quarks. Using
unitarity one obtains
\begin{eqnarray}
B_2 & = & {\mathrm Im}\; V_{tb} V_{4b}^* V_{4b'} V_{tb'}^*
= - {\mathrm Im}\; V_{tb} X_{bb'} V_{tb'}^* = T_2 + T_3 \,, \nonumber \\
B_1 & = & {\mathrm Im}\; V_{cb} V_{4b}^* V_{4b'} V_{cb'}^*
= -{\mathrm Im}\; V_{cb} X_{bb'} V_{cb'}^* = T_1 - T_3 \,, \nonumber \\
B_3 & = & {\mathrm Im}\; V_{cb} V_{tb}^* V_{tb'} V_{cb'}^* = T_3 \,,
\label{ec:26}
\end{eqnarray}
where $T_{1-3}$ are defined in Eqs. (\ref{ec:1}) and vanish in
the three generation SM.

These results can be also reproduced using explicitly the CKM matrix.
Using the freedom one has in the chiral limit to make unitary transformations
in the ($d$,$s$) space, one may write the CKM matrix in the form
\begin{equation}
V=\left( \begin{array}{cccc}
c_1 & s_1 c_3 & s_1 s_3 c_5 & s_1 s_3 s_5 \\
-s_1 c_2 & c_1 c_2 c_3 + s_2 s_3 c_6 e^{i \delta_1} &
c_1 c_2 s_3 c_5 - s_2 c_3 c_5 c_6 e^{i \delta_1} &
c_1 c_2 s_3 s_5 - s_2 c_3 s_5 c_6 e^{i \delta_1} \\
& & +s_2 s_5 s_6 e^{i(\delta_1 + \delta_3)} &
-s_2 c_5 s_6 e^{i(\delta_1+\delta_3)} \\
-s_1 s_2 & c_1 s_2 c_3 - c_2 s_3 c_6 e^{i \delta_1} &
c_1 s_2 s_3 c_5 + c_2 c_3 c_5 c_6 e^{i \delta_1} &
c_1 s_2 s_3 s_5 + c_2 c_3 s_5 c_6 e^{i \delta_1} \\
&  & -c_2 s_5 s_6 e^{i(\delta_1+\delta_3)} &
+c_2 c_5 s_6 e^{i(\delta_1+\delta_3)} \\
0 & s_3 s_6  & -c_3 c_5 s_6 -s_5 c_6 e^{i\delta_3} &
-c_3 s_5 s_6 +c_5 c_6 e^{i\delta_3}
\end{array}
\right) \,.
\label{ec:27}
\end{equation}
The invariants $B_i$ can then be expressed in terms of mixing angles and the
two physical CP violating phases,
\begin{eqnarray}
B_2 & = & {\mathrm Im}\; V_{tb} V_{4b}^* V_{4b'} V_{tb'}^* = T_2 + T_3 = 
\nonumber \\
& = &  c_1 s_2 c_2 s_3 c_3 s_6^2 c_6 (c_5^2 - s_5^2) \sin \delta_1
+ s_3^2 c_3 s_5 c_5 s_6 c_6 (c_2^2 - c_1^2 s_2^2) \sin \delta_3 \nonumber \\
& & + c_1 s_2 c_2 s_3 s_5 c_5 s_6 (c_6^2 - c_3^2) \sin (\delta_1+\delta_3)
+ c_1 s_2 c_2 s_3 c_3^2 s_5 c_5 s_6 c_6^2 \sin (\delta_1-\delta_3) \,,
\nonumber \\
B_1 & = & {\mathrm Im}\; V_{cb} V_{4b}^* V_{4b'} V_{cb'}^* = T_1 - T_3 = 
\nonumber \\
& = &  c_1 s_2 c_2 s_3 c_3 s_6^2 c_6 (s_5^2 - c_5^2) \sin \delta_1
+ s_3^2 c_3 s_5 c_5 s_6 c_6 (s_2^2 - c_1^2 c_2^2) \sin \delta_3  \nonumber \\
& & + c_1 s_2 c_2 s_3 s_5 c_5 s_6 (c_3^2 - c_6^2) \sin (\delta_1+\delta_3)
- c_1 s_2 c_2 s_3 c_3^2 s_5 c_5 s_6 c_6^2 \sin (\delta_1-\delta_3) \,,
\nonumber \\
B_3 & = & {\mathrm Im}\; V_{cb} V_{tb}^* V_{tb'} V_{cb'}^* = T_3 = \nonumber \\
& = &  c_1 s_2 c_2 s_3 c_3 s_6^2 c_6 (c_5^2 - s_5^2) \sin \delta_1
+ c_1^2 s_3^2 c_3 s_5 c_5 s_6 c_6 (c_2^2 - s_2^2) \sin \delta_3 \nonumber \\
& & + c_1 s_2 c_2 s_3 s_5 c_5 s_6 (c_1^2 s_3^2 - s_6^2) \sin (\delta_1+\delta_3)
+ c_1 s_2 c_2 s_3 c_3^2 s_5 c_5 s_6 c_6^2 \sin (\delta_1-\delta_3) \,.
\label{ec:28}
\end{eqnarray}
The unitary matrices in Eqs. (\ref{ec:22},\ref{ec:27}) have been used to
rederive the relevant imaginary parts of invariant quartets in the chiral
limits. Obviously, they do not correspond to the actual CKM matrix in the
physical situation where the quarks are distinguished.

\section{Limits on CP violating effects from new quarks}

In this Section we revise the three generation SM for which CP violation
is summarized in
a unitarity triangle, extending this description to the unitarity quadrangles
for an extra vector-like (sequential) quark (family) and their restriction to
subtriangles in the chiral limit. Then we estimate the experimental bounds on
the three independent invariants $B_i$ characterizing CP violation in this
case and comment on the determination of the new CP violating effects.

\subsection{Triangles, quadrangles and the chiral limit}

Before considering physics beyond the SM, it is worthwhile reviewing the main
features of CP violation in the three generation SM.
The information on CP violation is conventionally summarized in this case
in terms
of the unitarity triangle \cite{papiro14}.
This triangle is the geometrical representation
of the unitarity relations between any two different rows or columns of the
$3 \times 3$ CKM matrix. One can draw different triangles choosing different
pairs of rows or columns, but for the three generation SM unitarity implies that
all these triangles have the same area (which equals $|{\mathrm Im}\;
V_{ij} V_{kj}^* V_{kl} V_{il}^*|/2$) because all
${\mathrm Im}\;V_{ij} V_{kj}^* V_{kl} V_{il}^*$ have the same modulus, and 
this modulus gives the
strength of CP violation in the SM. The most interesting of the unitarity
triangles is the one which results from the orthogonality of the first and
third columns,
\begin{equation}
V_{ud} V_{ub}^* + V_{cd} V_{cb}^* + V_{td} V_{tb}^*=0\,.
\label{ec:29}
\end{equation}
The measurement of CP asymmetries in ${\mathrm B}^0$ decays offers the
possibility of measuring the internal angles of this triangle. Note that these
angles are rephasing invariant quantities since they are the arguments of
invariant quartets. In the three generation
SM, $|{\mathrm Im}\;V_{ij} V_{kj}^* V_{kl} V_{il}^*|$
is necessarily small, due
essentially to the fact that the third generation almost decouples from the
other two (in the limit where the third generation decouples there is no CP
violation in the SM). An upper limit on $|{\mathrm Im}\;V_{ij} V_{kj}^*
V_{kl} V_{il}^*|$ is readily obtained
since $|{\mathrm Im}\; V_{us} V_{cs}^* V_{cb} V_{ub}^*| \leq
|V_{us}|\,|V_{cs}|\,|V_{cb}|\,|V_{ub}| \leq 5 \times 10^{-5}$.

In the presence of an extra quark, the unitarity relation corresponding to Eq.
(\ref{ec:29}) becomes
\begin{equation}
V_{ud} V_{ub}^* + V_{cd} V_{cb}^* + V_{td} V_{tb}^* + V_{4d} V_{4b}^* =0\,.
\label{ec:30}
\end{equation}
In this case, three quadrangles are required to summarize the information on CP
violation \cite{papiro8}. The other two quadrangles can be chosen to be the
ones obtained multiplying the first and second columns and the second and third
columns, respectively.
The vanishing of their areas is a sufficient condition for CP
conservation, and in the case of nondegenerate quark masses it is also
necessary. Alternatively, one could choose a set of three quadrangles arising
from the orthogonality between the
rows of $V$. A relevant observation is that some of the sides of these
quadrangles cannot be measured separately in the case of degenerate quark
masses. For example,
in the limit where $m_{u,c}=0$ the quantities $V_{ud} V_{ub}^*$ and 
$V_{cd} V_{cb}^*$ in Eq. (\ref{ec:30}) cannot be separately measured, due to
the freedom to redefine degenerate quark fields. In this case, however, the
subtriangle with sides $V_{ud} V_{ub}^* + V_{cd} V_{cb}^*$, $V_{td} V_{tb}^*$
and $V_{4d} V_{4b}^*$ is well-defined.
In the chiral limits
previously considered, CP violation can be geometrically described as follows.

\begin{figure}
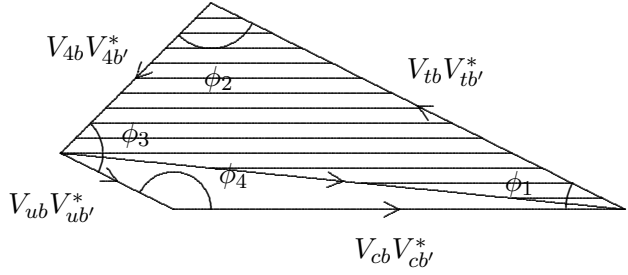

\begin{center}
\mbox{\beginpicture
\setcoordinatesystem units <1cm,1cm>
\unitlength=1cm
\linethickness=1pt
\setplotsymbol ({\thinlinefont .})
\plot 0 0 1.5 -0.75 /
\plot 1.5 -0.75 7.5 -0.75 /
\plot 7.5 -0.75 2 2 /
\plot 2 2 0 0 /
\plot 0 0 7.5 -0.75 /
\plot 0.5264 -0.375 0.75 -0.375 0.6158 -0.1961 /
\plot 4.3 -0.65 4.5 -0.75 4.3 -0.85 /
\plot 4.95 0.625 4.75 0.625 4.884 0.447 /
\plot 1.0707 1.212 1 1 1.212 1.0707 /
\plot 3.5609 -0.2556 3.75 -0.375 3.5410 -0.4546 /
\circulararc 71.57 degrees from 0.5 -0.25 center at 0 0
\circulararc 153.43 degrees from 2.0 -0.75 center at 1.5 -0.75
\circulararc 108.43 degrees from 1.57 1.57 center at 2 2
\circulararc 26.56 degrees from 6.8 -0.4 center at 7.5 -0.75
\linethickness=0.25pt
\plot 6 -0.6 7.2 -0.6 /
\plot 4 -0.4 6.8 -0.4 /
\plot 2 -0.2 6.4 -0.2 /
\plot 0 0 6 0 /
\plot 0.2 0.2 5.6 0.2 /
\plot 0.4 0.4 5.2 0.4 /
\plot 0.6 0.6 4.8 0.6 /
\plot 0.8 0.8 4.4 0.8 /
\plot 1 1 4 1 /
\plot 1.2 1.2 3.6 1.2 /
\plot 1.4 1.4 3.2 1.4 /
\plot 1.6 1.6 2.8 1.6 /
\plot 1.8 1.8 2.4 1.8 /
\put{$V_{ub} V_{ub'}^*$} [lB] at -0.7  -0.8
\put{$V_{cb} V_{cb'}^*$} [lB] at 3.9 -1.4
\put{$V_{tb} V_{tb'}^*$} [lB] at 4.6 1.0
\put{$V_{4b} V_{4b'}^*$} [lB] at -0.2 1.3
\put{$\phi_3$} [lB] at 0.8 0.15
\put{$\phi_2$} [lB] at 1.9 0.9
\put{$\phi_1$} [lB] at 5.9 -0.5
\put{$\phi_4$} [lB] at 2.1 -0.4
\linethickness=0pt
\putrectangle corners at -1 -2.5 and 8.5 3
\endpicture}
\end{center}
\caption{The shadowed triangle describes CP violation in the $m_{u,d,s,c}=0$
limit, whereas the complete quadrangle does it in the chiral limit
$m_{u,d,s}=0$. We use the same notation $\phi_{1,3}$ for the angles of the
shadowed
triangle as for the quadrangle, although for the (convex) quadrangle
 they are larger.}
\label{fig:1}
\end{figure}

In the $m_{u,d,s,c}=0$ limit, we can define the
subtriangle obtained multiplying the third and fourth columns of $V$
(shadowed region in Fig. \ref{fig:1}),
\begin{equation}
V_{ub} V_{ub'}^* +V_{cb} V_{cb'}^* +V_{tb} V_{tb'}^* +V_{4b} V_{4b'}^* = 0 \,,
\label{ec:31}
\end{equation}
and considering the sides
$V_{ub} V_{ub'}^* + V_{cb} V_{cb'}^*$, $V_{tb} V_{tb'}^*$,
$V_{4b} V_{4b'}^*$, with angles $\phi_{1-3}$,
\begin{eqnarray}
\sin \phi_1 & = & | \sin \arg (V_{ub} V_{tb}^* V_{tb'} V_{ub'}^* + V_{cb} 
V_{tb}^* V_{tb'} V_{cb'}^*) | \,, \nonumber \\
\sin \phi_2 & = & | \sin \arg V_{tb} V_{4b}^* V_{4b'} V_{tb'}^* | \,,
\nonumber \\
\sin \phi_3 & = & | \sin \arg (V_{ub} V_{4b}^* V_{4b'} V_{ub'}^* + V_{cb}
V_{4b}^* V_{4b'} V_{cb'}^* ) | \,.
\label{ec:32}
\end{eqnarray}
The area of this triangle represents the
strength of CP violation in this limit. This area is given by
\begin{equation}
A_{bb'} = \frac{1}{2} |B_2| = \frac{1}{2} | {\mathrm Im}\; V_{tb} V_{4b}^* 
V_{4b'} V_{tb'}^* | 
\label{ec:33}
\end{equation}
and vanishes if and only if CP is conserved (see Eqs. (\ref{ec:18},\ref{ec:19})
). Note that under allowed quark mass eigenstate transformations,
including those mixing $u$ and $c$, the length of the sides of the triangle
remains constant and therefore they are measurable quantities even in the limit
$m_{u,c}=0$, where $u$ and $c$ are indistinguishable.Thus this triangle provides
a good description of CP violation in the $m_{u,d,s,c}=0$ limit. The
subtriangle obtained multiplying the third and fourth rows,
\begin{equation}
V_{td} V_{4d}^* + V_{ts} V_{4s}^* + V_{tb} V_{4b}^* + V_{tb'} V_{4b'}^* = 0 \,,
\label{ec:34}
\end{equation}
with sides $V_{td} V_{4d}^* + V_{ts} V_{4s}^*$, $V_{tb} V_{4b}^*$,
$V_{tb'} V_{4b'}^*$ has the same
area $A_{t4} = | {\mathrm Im}\; V_{tb} V_{4b}^* V_{4b'} V_{tb'}^* |/2$
and provides an equivalent description of CP violation in this limit.

In the chiral limit, $m_{u,d,s}=0$, we have to consider the complete quadrangle
in Eq. (\ref{ec:31}), with sides $V_{ub} V_{ub'}^*$,
$V_{cb} V_{cb'}^*$, $V_{tb} V_{tb'}^*$, $V_{4b} V_{4b'}^*$
(see Fig. \ref{fig:1}) and angles
\begin{eqnarray}
\sin \phi_1 & = & |\sin \arg V_{cb} V_{tb}^* V_{tb'} V_{cb'}^* | \,,
\nonumber \\
\sin \phi_2 & = & |\sin \arg V_{tb} V_{4b}^* V_{4b'} V_{tb'}^* | \,,
\nonumber \\
\sin \phi_3 & = & |\sin \arg V_{ub} V_{4b}^* V_{4b'} V_{ub'}^* | \,,
\nonumber \\
\sin \phi_4 & = & |\sin \arg V_{ub} V_{cb}^* V_{cb'} V_{ub'}^* | \,. 
\label{ec:35}
\end{eqnarray}
The area of this quadrangle is
\begin{eqnarray}
A_{bb'}& = & \frac{1}{4} \{ |{\mathrm Im}\; V_{ub} V_{cb}^* V_{cb'} V_{ub'}^*| +
|{\mathrm Im}\; V_{cb} V_{tb}^* V_{tb'} V_{cb'}^*| \nonumber \\
& & + |{\mathrm Im}\; V_{tb} V_{4b}^* V_{4b'} V_{tb'}^*| +
|{\mathrm Im}\; V_{ub} V_{4b}^* V_{4b'} V_{ub'}^*|\} \nonumber \\
& = & \frac{1}{4} \{ |B_1+B_3| + |B_3| + |B_2| + |B_1+B_2| \} \,.
\label{ec:36}
\end{eqnarray}
We use the same notation as for the angles and area of the triangle in Eqs.
(\ref{ec:32},\ref{ec:33}) because this quadrangle reduces to that subtriangle
in the appropriate limit. It is clear that the vanishing of $A_{bb'}$ in Eq.
(\ref{ec:36}) is a necessary and sufficient condition for CP conservation.
Alternatively, one can consider adding to the triangle in Eq.
(\ref{ec:34}) the analogous triangles obtained multiplying the second and third
rows and the second and fourth rows respectively, with areas $A_{ct}=|B_3|/2 =
|{\mathrm Im}\;
V_{cb} V_{tb}^* V_{tb'} V_{cb'}^* |/2$ and $A_{c4}=|B_1|/2 = 
|{\mathrm Im}\;
V_{cb} V_{4b}^* V_{4b'} V_{cb'}^* |/2$. The vanishing of
$A_{ct,c4,t4}$ is also a necessary and sufficient condition for CP conservation.

\subsection{Bounds on $B_1$, $B_2$ and $B_3$}

We turn now to the important question of estimating the possible size of the
new CP violating effects when new quarks are added to the SM. In order to
establish upper bounds on the size of these effects one has to distinguish
between the case of an extra isosinglet quark and the case of a sequential
fourth family. In both cases, we use the experimental model-independent
measurements
\cite{papiro2} $|V_{ud}| = 0.9736 \pm 0.0010$, $|V_{us}| = 0.2205 \pm 0.0018$,
$|V_{cd}| = 0.224 \pm 0.016$, $|V_{cs}| = 1.01 \pm 0.18$, $|V_{ub}/V_{cb}| =
0.08 \pm 0.02$, $|V_{cb}| = 0.041 \pm 0.003$. These, together with the
unitarity of the $4 \times 4$ matrix $V$, give $|V_{ub'}| \leq 0.079$,
$|V_{cb'}| \leq 0.516$, $|V_{td,4d}| \leq 0.104$, $|V_{ts,4s}| \leq 0.513$,
where we have used the measured lower bounds to obtain these upper limits. Our
strategy will be to obtain rigorous upper bounds on $|B_i|$ using the previous
limits and to check
afterwards that they are almost saturated in particular cases fulfilling all
present experimental constraints. 
The former upper bounds imply
$|{\mathrm Im}\; V_{ub} V_{cb}^* V_{cb'} V_{ub'}^*| \leq |V_{ub}|\, |V_{cb}|
\, |V_{cb'}| \, |V_{ub'}| \leq 7.87 \times 10^{-6}$,
$|{\mathrm Im}\; V_{ub} V_{tb}^* V_{tb'} V_{ub'}^*| \leq |V_{ub}|\, |V_{tb}|
\, |V_{tb'}| \, |V_{ub'}| \leq 1.73 \times 10^{-4}$,
$|{\mathrm Im}\; V_{cb} V_{tb}^* V_{tb'} V_{cb'}^*| \leq |V_{cb}|\, |V_{tb}|
\, |V_{tb'}| \, |V_{cb'}| \leq 1.11 \times 10^{-2}$. Then, using unitarity and
the triangular inequality for the absolute values, these limits translate into
\begin{eqnarray}
|B_1| & = & |{\mathrm Im}\; V_{ub} V_{cb}^* V_{cb'} V_{ub'}^* -
{\mathrm Im}\; V_{cb} V_{tb}^* V_{tb'} V_{cb'}^*| \leq 1.11 \times 10^{-2} \,,
\nonumber \\
|B_2| & = & |{\mathrm Im}\; V_{ub} V_{tb}^* V_{tb'} V_{ub'}^* +
{\mathrm Im}\; V_{cb} V_{tb}^* V_{tb'} V_{cb'}^*| \leq 1.12 \times 10^{-2} \,,
\nonumber \\
|B_3| & = & |{\mathrm Im}\; V_{cb} V_{tb}^* V_{tb'} V_{cb'}^*| \leq 1.11 \times
10^{-2} \,,
\label{ec:n1}
\end{eqnarray}
which are rigorous bounds, in particular for a fourth family.
These bounds are mostly saturated for instance
by the $4 \times 4$ unitary matrix
\begin{equation}
|V| = \left( \begin{array}{cccc}
0.973 & 0.220 & 0.0035 & 0.070 \\
0.230 & 0.918 & 0.041 & 0.321 \\
0.082 & 0.254 & 0.655 & 0.712 \\
0.082 & 0.212 & 0.755 & 0.621
\end{array} \right) \,,
\label{ec:n4}
\end{equation}
\begin{equation}
\arg V = \left( \begin{array}{cccc}
0 & 0 & 0 & 0 \\
\pi & 0.007 & -1.08 & -0.057 \\
\pi & 0.095 & 0.873 & -3.00 \\
\pi & -1.31 & 2.38 & 1.62
\end{array} \right) \,,
\label{ec:n4b}
\end{equation}
for which $|B_1|=6.1 \times 10^{-3}$, $|B_2|=6.3 \times 10^{-3}$,
$|B_3|=6.1 \times 10^{-3}$. We have also required in these matrices
that the imaginary parts of the quartets involving the first two columns and
entering in the calculation of
$\epsilon_{\mathrm K}$ for four sequential families
are $\sim 10^{-4}$ not to rely on large cancellations.
Without this requirement the bounds in Eqs. (\ref{ec:n1}) can be almost
completely saturated. Hence $|B_i| \leq 10^{-2}$ for four families.

In the case of an extra isosinglet quark the size of the CKM matrix elements is
further constrained by existing bounds on FCNC \cite{papiro11}. 
The constraint on $|V_{4d}|\, |V_{4s}| = |X_{ds}|$ is rather severe due to
the experimental upper bound on strangeness changing neutral currents. The
strongest limit on $|X_{ds}|$ arises from the experimental bound
\cite{papiro2}
\begin{equation}
\mathrm Br\, ( K^+ \rightarrow \pi^+ \nu \bar \nu)=\frac{\Gamma \,
( K^+ \rightarrow \pi^+ \nu \bar \nu)}{\Gamma (K^+ \rightarrow all)} < 2.4
\times 10^{-9}\,.
\label{ec:g3}
\end{equation}
Comparison with the process $\mathrm K^+ \rightarrow \pi^0 e^+ \nu$ leads to
\begin{equation}
\frac{\mathrm Br\, ( K^+ \rightarrow \pi^+ \nu \bar \nu)}
{\mathrm Br\, ( K^+ \rightarrow \pi^0 e^+ \nu)}=\frac{|X_{ds}|^2}{2 |V_{us}|^2} \times
3\,,
\label{ec:g4}
\end{equation}
where the factor 3 takes into account the three different $\nu \bar \nu$ pairs.
Then the observed
$\mathrm Br\, ( K^+ \rightarrow \pi^0 e^+ \nu)=(4.82 \pm 0.06)\%$ \cite{papiro2}
gives $|X_{ds}| < 4.08 \times 10^{-5}$. 
The limits on $|X_{db}| = |V_{4d}| \, |V_{4b}|$,
$|X_{sb}| = |V_{4s}| \, |V_{4b}|$ arise from the experimental bound
\cite{papiro2}
\begin{equation}
\mathrm \frac{\Gamma \, (B \rightarrow \mu^+ \mu^- X)}
{\Gamma \, (B \rightarrow \mu \nu X)} < 4.6 \times 10^{-4}\,,
\label{ec:g5}
\end{equation}
which leads to \cite{papiro11}
\begin{equation}
\frac{|X_{db}|^2 + |X_{sb}|^2}{|V_{ub}|^2 + 
R\, |V_{cb}|^2} < 3.67 \times 10^{-3} \,,
\label{ec:g6}
\end{equation}
where $R \simeq 0.5$ is a phase space factor, giving $|X_{db,sb}| \leq 1.91
\times 10^{-3}$. Using this limit, the experimental bounds above and
the triangular inequality for the absolute values we obtain
\begin{eqnarray}
|{\mathrm Im}\; V_{ub} V_{4b}^* V_{4b'} V_{ub'}^*| & = & 
|{\mathrm Im}\; V_{ud} V_{4d}^* V_{4b} V_{ub}^* +  
{\mathrm Im}\; V_{us} V_{4s}^* V_{4b} V_{ub}^*| \leq 1.01 \times 10^{-5} \,,
\nonumber \\
|{\mathrm Im}\; V_{cb} V_{4b}^* V_{4b'} V_{cb'}^*| & = & 
|{\mathrm Im}\; V_{cd} V_{4d}^* V_{4b} V_{cb}^* +  
{\mathrm Im}\; V_{cs} V_{4s}^* V_{4b} V_{cb}^*| \leq 1.02 \times 10^{-4} \,,
\label{ec:n2}
\end{eqnarray}
which together with the general bound
$|{\mathrm Im}\; V_{ub} V_{cb}^* V_{cb'} V_{ub'}^*| \leq 7.87 \times 10^{-6}$
translate into
\begin{eqnarray}
|B_1| & = & |{\mathrm Im}\; V_{cb} V_{4b}^* V_{4b'} V_{cb'}^*| \leq 1.02 \times
10^{-4} \,, \nonumber \\
|B_2| & = & |{\mathrm Im}\; V_{ub} V_{4b}^* V_{4b'} V_{ub'}^* +
{\mathrm Im}\; V_{cb} V_{4b}^* V_{4b'} V_{cb'}^*| \leq 1.12 \times 10^{-4} \,,
\nonumber \\
|B_3| & = & |{\mathrm Im}\; V_{ub} V_{cb}^* V_{cb'} V_{ub'}^* +
{\mathrm Im}\; V_{cb} V_{4b}^* V_{4b'} V_{cb'}^*| \leq 1.11 \times 10^{-4} \,.
\label{ec:n3}
\end{eqnarray}
These are the rigorous limits for an extra down quark isosinglet. (We have not
made explicit use of the $|X_{ds}|$ bound.) The $4 \times 4$ unitary matrix
\begin{equation}
|V|= \left( \begin{array}{cccc}
0.975 & 0.222 & 0.0033 & 0.0007 \\
0.222 & 0.974 & 0.039 & 0.013 \\
0.011 & 0.039 & 0.978 & 0.204  \\
0.0022 & 0.0093 & 0.204 & 0.979  
\end{array} \right) \,,
\label{ec:n5}
\end{equation}
\begin{equation}
\arg V= \left( \begin{array}{cccc}
0 & 0 & 0 & 0 \\
\pi & 0.0006 & -1.97 & -2.09 \\
\pi & -0.300 & 0.832 & 1.59 \\
\pi & -0.234 & 2.56 & 0.142 
\end{array} \right)
\label{ec:n5b}
\end{equation}
gives $|B_1|=7.6 \times 10^{-5}$, $|B_2|=7.7 \times 10^{-5}$, $|B_3| = 7.6
\times 10^{-5}$ which are near the upper bounds in Eqs. (\ref{ec:n3}).
We also require that
the dominant contributions to $\epsilon_{\mathrm K}$ are
the same as in the three generation SM without large cancellations,
and not mediated by $Z$ tree level
diagrams \cite{papiro6,papiro11}.

\subsection{CP violation from new quarks}

Vector-like and sequential quark contributions to CP violating observables not
distinguishing between $d$ and $s$ quarks are proportional to $B_i$. We have
shown that for vector-like and sequential quarks $|B_i| \leq 10^{-4}$ and
$|B_i| \leq 10^{-2}$, respectively.
These values are relatively large. For instance, the
maximum of $|{\mathrm Im}\;V_{ij} V_{kj}^* V_{kl} V_{il}^*|$ 
for an arbitrary $4 \times 4$ unitary matrix is $1/6 \sqrt
3 \simeq 0.096$, which is the same as for a $3 \times 3$ unitary matrix
\cite{papiro16}. On the other hand, 
$|{\mathrm Im}\;V_{ij} V_{kj}^* V_{kl} V_{il}^*| \leq 5 \times 10^{-5}$ in the
three generation SM. In spite of the relatively large values allowed for $B_i$,
it is clear that observing direct CP violation from gauge couplings of new
quarks will not be an easy task. It is worth emphasizing that $B_i$ can also be
obtained indirectly, by measuring $T_i$ and using Eqs. (\ref{ec:26}) which give
$B_i$ as functions of $T_i$.
The failure of the three
generation SM unitarity relations would point out to new (CP violating)
physics, in particular to new quarks if $B_i$ in Eqs. (\ref{ec:26}) are of the
correct size. The study of CP violation at high energies would thus complement
the information of CP asymmetries in B meson decays, at B factories. The effects
of vector-like or sequential quarks on the CP asymmetries in ${\mathrm B}^0$
decays have been extensively studied in the literature \cite{papiro11}.  In
the case of vector-like quarks, the most important effect results from a new
contribution to ${\mathrm B}_d-\bar{\mathrm B}_d$ and 
${\mathrm B}_s-\bar{\mathrm B}_s$ mixings, arising from tree level FCNC
Z exchange diagrams. It has been shown \cite{papiro11} that even for
relatively small FCNC couplings, the prediction for the CP asymmetries in
${\mathrm B}_d^0 \rightarrow {\mathrm J}/\psi\;{\mathrm K}_0$ and ${\mathrm
B}_d^0 \rightarrow \pi^+\,\pi^-$ decays can differ significantly from the
predictions of the SM. At this point, it should be emphasized that although the
observation of CP asymmetries at B factories may lead to unambiguous evidence
for new physics, it will not be easy to identify the origin of the new physics
by studying CP asymmetries alone \cite{papironn2}. The study of CP violation at
high energies, together with the study of rare B decays, will play an important
r\^{o}le in identifying the origin of the new sources of CP violation.

\section{Conclusions}

Understanding the origin of CP violation will probably require obtaining new
experimental information on CP violating observables outside the kaon system and
the possible identification of new sources of CP violation. One of the simplest
ways of obtaining new sources of CP violation consists of adding extra quarks to
the SM. The addition of extra vector-like fermions is specially attractive,
since they naturally arise in grand-unified theories, like ${\mathrm E}_6$. We
have derived a complete set of weak-basis invariants which constitute necessary
and sufficient conditions for CP invariance. These weak-basis invariants are
physical quantities which can be expressed in terms of quark masses and
imaginary parts of rephasing invariant quartets. For simplicity we have
restricted ourselves to the case of one additional isosinglet quark, since it is
sufficient to illustrate the implications of new quarks for observables
involving only known fermions. 

At this point, it is worth emphasizing the usefulness of weak basis invariants
in the study of CP violation:
\begin{enumerate}
\item  The invariant approach can be very useful in
model building. At the moment, there is no standard theory of flavour and for
example in the SM the Yukawa couplings are arbitrary free parameters. As a
result, one does not have in the SM any insight into the pattern of fermion
masses and mixings. In the literature, there have been various attempts of
introducing additional family symmetries in the Lagrangian leading to Yukawa
couplings which are no longer arbitrary but  are expressed in terms of a fewer
number of parameters, with the Yukawa couplings exhibiting some texture zeros
\cite{papironn3}. Of course the quark mass matrices are no longer arbitrary,
being constrained by the family symmetries. One has to check whether in
spite of the additional family symmetries the model leads to genuine CP
violation mediated by W interactions. The usual method of diagonalizing the
quark mass matrices becomes rather inadequate, specially in models with
vector-like quarks, where the CKM matrix is no longer a unitary matrix. The
simplest way of checking whether CP violation occurs in models with additional
family symmetries consists of directly evaluating the weak-basis invariants
which constitute the necessary and sufficient conditions for CP invariance in
the model considered. If any of these invariants is non-vanishing one is sure to
have CP violation.

\item Weak-basis invariants are also very useful for studying CP violation in
various physical limits,
especially those involving degenerate and vanishing masses. Here we
discuss two chiral limits, the extreme one $m_t \gg m_c \sim 0$, $m_{u,d,s}=0$
and the standard chiral limit $m_{u,d,s}=0$. These limits are specially relevant
at high energy colliders, where the natural asymptotic states are quark jets. In
this chiral limit, $d$ and $s$ quark jets are either very difficult or
impossible to distinguish from each other and there are no CP violation effects
in the three generation SM. In the case of one extra vector-like quark or an
extra sequential family we have shown, using weak-basis invariants, that there
is CP violation even in this chiral limit. We have shown that CP violation can
be characterized by two CP violating phases which are proportional to $B_1=
{\mathrm Im}\; V_{cb} V_{4b}^* V_{4b'} V_{cb'}^*$, $B_2=
{\mathrm Im}\; V_{tb} V_{4b}^* V_{4b'} V_{tb'}^*$ and $B_3=
{\mathrm Im}\; V_{cb} V_{tb}^* V_{tb'} V_{cb'}^*$. In the extreme chiral limit
($m_{u,d,s,c}=0$) we have shown that there is one CP violating phase and one
weak-basis invariant which controls the strength of CP violation in this limit
and is proportional to $B_2$. 
\end{enumerate}

In conclusion, extra quarks lead to new sources of CP violation which can
manifest themselves in various phenomena, including CP asymmetries in ${\mathrm
B}^0$ decays, rare B decays as well as in CP violating observables at high
energy. Weak-basis invariants, together with the imaginary part of rephasing
invariant quartets, like $B_i$ and $T_i$ which we have introduced, are useful
tools to study CP violation in these minimal extensions of the SM.

\vspace{1cm}
\noindent
{\Large \bf Acknowledgements}

\vspace{0.4cm}
\noindent
We thank J. Bernab\'{e}u and 
M. Zra{\l}ek for discussions. This work was partially supported by
CICYT under contract AEN96-1672, by the Junta de Andaluc\'{\i}a and by the
European Union under contract CHRX-CT92-0004.

\appendix
\section{Appendix}
The Lagrangian 
${\mathcal L}_{\mathrm gauge}+{\mathcal L}_{\mathrm mass}$ 
in Eqs. (\ref{ec:4a},\ref{ec:4b})
is invariant under CP
if and only if there exist unitary transformations $U_L$, $U_L^{d}$,
$U_R^{u,d}$ such that \cite{papiro4b}
\begin{equation}
q_{L}^{(d)} \rightarrow U_L C q_{L}^{(d)*} ~,~~
d_{L}^{(s)} \rightarrow U_L^d C d_{L}^{(s)*} ~,~~
q_{R}^{(s)} \rightarrow U_R^q C q_{R}^{(s)*} \,,
\label{ec:ap1}
\end{equation}
with $C$ the Dirac charge-conjugation matrix, satisfying
\begin{equation}
U_L^\dagger M_q U_R^q = M_q^* ~,~~
U_L^{d\dagger} m_d U_R^d = m_d^*\,.
\label{ec:ap2}
\end{equation}
As $U_R^{u,d}$ are unobservable (if Higgs mediated interactions are neglected)
we can assume ${\mathcal M}_u$ and ${\mathcal M}_d$ in Eq. (\ref{ec:5}) 
hermitian with nonnegative eigenvalues. Then,
\begin{equation}
U_L^\dagger H_u U_L=H_u^* ~,~~
U_L^\dagger H_d U_L=H_d^* ~,~~
U_L^\dagger h_d U_L^d\;=\;h_d^* ~,~~
U_L^{d \dagger} h'_d U_L^d={h'}_d^* \,,
\label{ec:ap5}
\end{equation}
with ${h'}_d=m_d m_d^\dagger$,
are equivalent to Eqs. (\ref{ec:ap2}) and are also
necessary and sufficient conditions for CP conservation. (The condition
$U_L^{d \dagger} h_d^\dagger U_L=h_d^T$ follows trivially from Eq.
(\ref{ec:ap5}). )

From these equalities new constraints for CP invariance can be derived which are
independent of the choice of weak basis, with no reference to unitary matrices
as in Eqs. (\ref{ec:ap1},\ref{ec:ap2},\ref{ec:ap5}). They result from the
observation that any combination of products of $H_u$, $H_d$, $h_d {h'}_d^p
h_d^\dagger$, with $p$ arbitrary, has invariant trace and determinant. Then CP
invariance, Eq. (\ref{ec:ap5}), requires that their imaginary part vanishes.
This
also holds for any combination of products of $h'_d$ and $h_d^\dagger H h_d$,
with $H$ any of the former combinations, but there is no need to consider these
combinations because they do not give new constraints. Which subsets of these
constraints are also sufficient has to be determined case by case. The explicit
proof can be done in a simple, convenient basis. For the search of necessary
and sufficient constraints and in general for parametrizing the model, we find
convenient to consider the basis where ${\mathcal M}_u=M_u$ is diagonal with
nonnegative eigenvalues and
\begin{equation}
{\mathcal M}_d= \left( \begin{array}{c}
M_d \\  m_d \end{array} \right) =
\left( \begin{array}{cc}
\tilde M_d & N_d \\
0 & \tilde m_d \end{array} \right)\,,
\label{ec:ap6}
\end{equation}
with $\tilde M_d$ upper triangular with real, nonnegative diagonal elements,
$\tilde m_d$ diagonal with nonnegative eigenvalues and $N_d$ arbitrary ($\tilde
M_d$ could also be chosen to be hermitian).

The proliferation of invariant constraints required to
guarantee CP invariance makes necessary the use of a symbolic algebraic program
to write down the expressions and to solve explicitly the
constraints. This is done with {\em Mathematica} \cite{papiroa1} 
and a set of routines analogous to those in Ref. \cite{papiroa2}.

$N=3$, $n_d=1$. In this case
we shall show that $I_{1-7}=0$ in Eqs. (\ref{ec:16}) is a
set of necessary and sufficient conditions for CP conservation.
In the proof we assume $M_u$ diagonal with
$(M_u)_{ij}=m_i \delta_{ij}$ and ${\mathcal M}_d$ upper triangular with
matrix elements $({\mathcal M}_d)_{i \leq j}=n_{ij}$. We consider the products
of $H_{u,d}$ and $h_d$ in Eq. (\ref{ec:ap5}), ordering them by increasing
number of factors. Then
the imaginary part of such products give the invariant conditions we
look for.

The lowest order invariant not identically zero $I_1$ has 8 mass submatrix
factors and gives in the convenient basis of Eq. (\ref{ec:ap6}) the condition
\begin{eqnarray}
I_1 & = & (m_1^2-m_2^2) |n_{44}|^2 ( {\mathrm Im}\;n_{12} n_{22}^* n_{24}
n_{14}^* + {\mathrm Im}\;n_{13} n_{23}^* n_{24} n_{14}^*) \nonumber \\
& & + (m_1^2-m_3^2) |n_{44}|^2 {\mathrm Im}\;n_{13} n_{33}^* n_{34} n_{14}^*
+ (m_2^2-m_3^2) |n_{44}|^2 {\mathrm Im}\;n_{23} n_{33}^* n_{34} n_{24}^*=0\,.
\label{ec:ap8}
\end{eqnarray}
The expression of $I_1$ suggests that we consider products with higher powers
of $H_u$, to obtain independent linear combinations of the imaginary factors.
In this way we find
\begin{eqnarray}
I_2 & = & (m_1^2+m_2^2) (m_1^2-m_2^2) |n_{44}|^2 ( {\mathrm Im}\;n_{12} 
n_{22}^* n_{24} n_{14}^* + {\mathrm Im}\;n_{13} n_{23}^* n_{24} n_{14}^*) 
\nonumber \\
& & + (m_1^2+m_3^2) (m_1^2-m_3^2) |n_{44}|^2 {\mathrm Im}\;n_{13} n_{33}^* 
n_{34} n_{14}^* \nonumber \\
& & + (m_2^2+m_3^2) (m_2^2-m_3^2) |n_{44}|^2 {\mathrm Im}\;n_{23} n_{33}^* 
n_{34} n_{24}^*=0\,, \nonumber \\
I_3 & = & (m_1^4+m_2^4) (m_1^2-m_2^2) |n_{44}|^2 ( {\mathrm Im}\;n_{12} 
n_{22}^* n_{24} n_{14}^* + {\mathrm Im}\;n_{13} n_{23}^* n_{24} n_{14}^*) 
\nonumber \\
& & + (m_1^4+m_3^4) (m_1^2-m_3^2) |n_{44}|^2 {\mathrm Im}\;n_{13} n_{33}^* 
n_{34} n_{14}^* \nonumber \\
& & + (m_2^4+m_3^4) (m_2^2-m_3^2) |n_{44}|^2 {\mathrm Im}\;n_{23} n_{33}^* 
n_{34} n_{24}^*=0\,.
\label{ec:ap9}
\end{eqnarray}
We will assume for the moment $n_{44} \neq 0$ and nondegenerate masses. Then
Eqs. (\ref{ec:ap8},\ref{ec:ap9}) imply
\begin{eqnarray}
{\mathrm Im}\;n_{12} n_{22}^* n_{24} n_{14}^* + 
{\mathrm Im}\;n_{13} n_{23}^* n_{24} n_{14}^* & = & 0 \,, \nonumber \\
{\mathrm Im}\;n_{13} n_{33}^* n_{34} n_{14}^* =
{\mathrm Im}\;n_{23} n_{33}^* n_{34} n_{24}^* & = & 0 \,.
\label{ec:ap10}
\end{eqnarray}
These conditions do not guarantee CP conservation, hence we go on considering
the
products with increasing number of factors and giving independent conditions.
The next lowest order invariant $I_4$ has 10 mass submatrix factors and after 
substituting (\ref{ec:ap10}) it can be written
\begin{eqnarray}
I_4 & = & (m_2^2-m_1^2) |n_{33}|^2 |n_{44}|^2 {\mathrm Im}\; n_{12} n_{22}^*
n_{24} n_{14}^*  \nonumber \\
& & +  (m_3^2-m_2^2) |n_{44}|^2 {\mathrm Im}\;n_{12} n_{22}^* n_{24} n_{34}^*
n_{33} n_{13}^* \nonumber \\
& & + (m_1^2-m_3^2) |n_{44}|^2 {\mathrm Im}\;n_{12} n_{22}^* n_{23} n_{33}^*
n_{34} n_{14}^*=0\,.
\label{ec:ap11}
\end{eqnarray}
We again look to products with higher powers of $H_u$ to obtain independent
linear
combinations of the imaginary factors, finding
\begin{eqnarray}
I_5 & = & (m_1^2+m_2^2) (m_2^2-m_1^2) |n_{33}|^2 |n_{44}|^2 {\mathrm Im}\; 
n_{12} n_{22}^* n_{24} n_{14}^* \nonumber \\
& & + (m_2^2+m_3^2) (m_3^2-m_2^2) |n_{44}|^2 {\mathrm Im}\;n_{12} n_{22}^* 
n_{24} n_{34}^* n_{33} n_{13}^* \nonumber \\
& & + (m_1^2+m_3^2) (m_1^2-m_3^2) |n_{44}|^2 {\mathrm Im}\;n_{12} n_{22}^* 
n_{23} n_{33}^* n_{34} n_{14}^*=0\,, \nonumber \\
I_6 & = & (m_1^4+m_2^4) (m_2^2-m_1^2) |n_{33}|^2 |n_{44}|^2 {\mathrm Im}\; 
n_{12} n_{22}^* n_{24} n_{14}^* \nonumber \\
& & + (m_2^4+m_3^4) (m_3^2-m_2^2) |n_{44}|^2 {\mathrm Im}\;n_{12} n_{22}^* 
n_{24} n_{34}^* n_{33} n_{13}^* \nonumber \\
& & + (m_1^4+m_3^4) (m_1^2-m_3^2) |n_{44}|^2 {\mathrm Im}\;n_{12} n_{22}^* 
n_{23} n_{33}^* n_{34} n_{14}^*=0\,.
\label{ec:ap12}
\end{eqnarray}
These equations imply for $n_{44} \neq 0$ and nondegenerate masses
\begin{equation}
|n_{33}|^2  {\mathrm Im}\; n_{12} n_{22}^* n_{24} n_{14}^* =
{\mathrm Im}\; n_{12} n_{22}^* n_{24} n_{34}^* n_{33} n_{13}^* = 
{\mathrm Im}\; n_{12} n_{22}^* n_{23} n_{33}^* n_{34} n_{14}^* = 0 \,.
\label{ec:ap13}
\end{equation}
A tedious calculation shows that Eqs. (\ref{ec:ap10},\ref{ec:ap13})
do imply CP conservation. First we find all the
solutions to Eqs. (\ref{ec:ap10},\ref{ec:ap13}) with all $n_{ij} \neq 0$,
then
with one $n_{ij}=0$, with two, etc.
In all cases we can redefine the quark eigenstate phases
to make ${\mathcal M}_d$ real.

When two up quark masses are degenerate, say $m_1=m_2$, we can assume without
loss of
generality $n_{12}=0$. Then, $I_{2,3}$ are proportional to $I_1$ and $I_{5,6}$
to $I_4$. Whereas
\begin{eqnarray}
I_1 & = & (m_1^2-m_3^2) |n_{44}|^2 ( {\mathrm Im}\; n_{13} n_{33}^* n_{34}
n_{14}^* + {\mathrm Im}\; n_{23} n_{33}^* n_{34} n_{24}^*) = 0\,, \nonumber \\
I_4 & = & (m_1^2-m_3^2) |n_{44}|^2 ( |n_{11}|^2 {\mathrm Im}\; n_{13} n_{33}^*
n_{34} n_{14}^*+|n_{22}|^2 {\mathrm Im}\; n_{23} n_{33}^* n_{34}n_{24}^*)=0\,.
\label{ec:ap14}
\end{eqnarray}
If $|n_{11}| \neq |n_{22}|$, these equations are independent and
\begin{equation}
{\mathrm Im}\; n_{13} n_{33}^* n_{34} n_{14}^* =
{\mathrm Im}\; n_{23} n_{33}^* n_{34} n_{24}^* = 0 \,.
\label{ec:ap15}
\end{equation}
If $|n_{11}|=|n_{22}|$, we can assume $n_{13}=0$ and Eqs.
(\ref{ec:ap15}) still hold. A long and tedious calculation shows that Eqs.
 (\ref{ec:ap15}) imply
CP conservation. We look for all their solutions and check that we can 
redefine the weak quark
basis conveniently and make ${\mathcal M}_d$ real for each solution. 
(In most cases it is only necessary to redefine the phases of the eigenstates.)
If the three up quark masses are degenerate, CP is conserved.

When $n_{44}=0$, $I_{1-6}=0$ and we need to introduce more constraints on
the mass matrices to ensure CP conservation. In this case
we can assume $n_{11}=n_{22}=n_{33}=0$
by properly choosing the weak basis. There is only one CP
violating phase, and the vanishing of the generalization of the SM invariant,
\begin{eqnarray}
I_7 & = & (m_2^2-m_1^2) (m_1^2-m_3^2) (m_3^2-m_2^2) |n_{34}|^2 {\mathrm Im}\;
n_{13} n_{23}^* n_{24} n_{14}^* =0
\label{ec:ap16}
\end{eqnarray}
does ensure CP conservation. What completes the proof.

$N=3$, $n_d>1$. In these cases the number of necessary and sufficient
invariant conditions for CP invariance is too large to be in general
manageable. Let us argue the fast growth of the number of these constraints by
deriving lower bounds for $n_d=2$ and $n_d=3$. These bounds are 
general and based on cycle
counting. A k-cycle is a product of $k$ matrix elements $n_{ij}$ of ${\mathcal
M}_d$:
${\mathcal C}(i_1,\dots,i_k)=\tilde n_{i_1
i_2} \tilde n_{i_2 i_3} \cdots \tilde n_{i_k i_1}$, where the indices $i_j$ are
all different and $\tilde n_{ij}$ can be $n_{ij}$ or $n_{ji}^*$.
The number $p_{\mathrm min}$ of invariant constraints obtained by this method is
smaller than the actual
number $p$ for (i) we consider only in this counting nondegenerate up masses,
and (ii) we assume that
only one invariant is needed to ensure the reality of a cycle (although we know
that often this is not the case, see Refs. \cite{papiro8,papiroa2} for
examples). Then comparing with the exact result for the simplest case we expect
$p_{\mathrm min} < p \sim 2 p_{\mathrm min}$.

For $n_d=2$ there are seven 3-cycles ${\mathcal
C}(1,2,3)$, ${\mathcal C}(1,2,4)$, ${\mathcal C}(1,2,5)$, ${\mathcal C}(1,3,4)$,
${\mathcal C}(1,3,5)$, ${\mathcal C}(2,3,5)$, ${\mathcal C}(2,3,4)$.
We work in a convenient basis where $n_{45}=n_{54}=0$, so the cycles with
$\tilde n_{45}$ are zero (for instance, the 3-cycles
${\mathcal C}(1,4,5)$, ${\mathcal C}(2,4,5)$ and ${\mathcal C}(3,4,5)$ ). 
To ensure the reality of these seven cycles we need seven constraints. In
addition, there are situations in which all the 3-cycles are real but not
necessarily the 4-cycles. This happens when some ${\mathcal M}_d$ matrix
elements vanish. The maximum number of nonreal 4-cycles is achieved
for instance if $n_{12}=n_{13}=n_{23}=0$. We have in this case three 4-cycles
not necessarily real ${\mathcal C}(1,4,2,5)$, ${\mathcal C}(1,4,3,5)$, 
${\mathcal C}(2,4,3,5)$, and
to ensure their reality we need three more constraints. Their reality
then implies the reality of the 5-cycles. Thus, $p_{\mathrm
min}=10$ for $n_d=2$. The analogous computation gives
$p_{\mathrm min}=20$ for $n_d=3$.
Finally if we perform the computation for $n_d=1$ we find $p_{\mathrm min}=4$
and we have shown that $p=7$. 
It must be noted that for $n_d=1$ the 3-cycle ${\mathrm C}(1,2,3)$
does not appear in the expressions of the invariants in this Appendix. It
should appear in $I_7$ but due to the basis redefinition it is replaced
by ${\mathcal C}(1,3,2,4)$.

\end{document}